\documentclass[12pt]{article}
\usepackage{scicite}
\usepackage{times}
\usepackage{graphicx}
\usepackage{url}
\usepackage{mathcomp,color}

\newenvironment{sciabstract}{%
\begin{quote} \bf}
{\end{quote}}

\newcounter{lastnote}

\title{Dimensional analysis identifies contrasting dynamics of past climate states and critical transitions}
\author{
\hskip-10exTommaso Alberti,$^{1\ast}$ Fabio Florindo,$^{1}$ Eelco J. Rohling,$^{2,3}$ Valerio Lucarini,$^{4,5}$ \\
\hskip-10exDavide Faranda$^{6,7,8}$
\\
\\
\normalsize{$^{1}$Istituto Nazionale di Geofisica e Vulcanologia, Rome 00143, Italy}
\\
\hskip-10ex\normalsize{$^{2}$Research School of Earth Sciences, Australian National University, Canberra, ACT 2601, Australia}
\\
\hskip-13ex\normalsize{$^{3}$School of Ocean and Earth Science, University of Southampton, National Oceanography Centre, Southampton SO14 3ZH, UK} 
\\
\hskip-10ex\normalsize{$^{4}$Department of Mathematics and Statistics, University of Reading, Reading RG6 6AH, UK} 
\\
\hskip-10ex\normalsize{$^{5}$Centre for the Mathematics of Planet Earth, University of Reading, Reading RG6 6AX, UK}
\\
\hskip-10ex\normalsize{$^{6}$Laboratoire des Sciences du Climat et de l’Environnement, Université Paris-Saclay, Gif-sur-Yvette 91191, France}
\\
\hskip-10ex\normalsize{$^{7}$London Mathematical Laboratory, London W6 8RH, UK}
\\
\hskip-10ex\normalsize{$^{8}$LMD/IPSL, Ecole Normale Superieure, PSL research University, Paris 75005, France}
\\
\hskip-15ex\normalsize{$^\ast$To whom correspondence should be addressed; E-mail: tommaso.alberti@ingv.it.}
}

\date{\today}

\begin{document} 
\baselineskip24pt
\maketitle

\begin{sciabstract}
While one can unequivocally identify past climate transitions, we lack comprehensive knowledge about their underlying mechanisms and timescales. Our study employs a dimensional analysis of benthic stable isotope records to uncover, across different timescales, how the climatic fluctuation of the Cenozoic are associated with changes in the number of effective degrees of freedom. Precession timescales dominate the Hothouse and Warmhouse states, while the Icehouse climate is primarily influenced by obliquity and eccentricity timescales. Notably, the Coolhouse state lacks dominant timescales. Our analysis proves effective in objectively identifying abrupt climate shifts and extremes. This is also demonstrated using high-resolution data from the last glacial cycle, revealing abrupt climate shifts within a single climate state. These findings significantly impact our understanding of the inherent stability of each climate state and the evaluation of (paleo-)climate models' ability to replicate key features of past/future climate states and transitions.
\end{sciabstract}

\begin{figure}[!h]
    \includegraphics[scale=0.5]{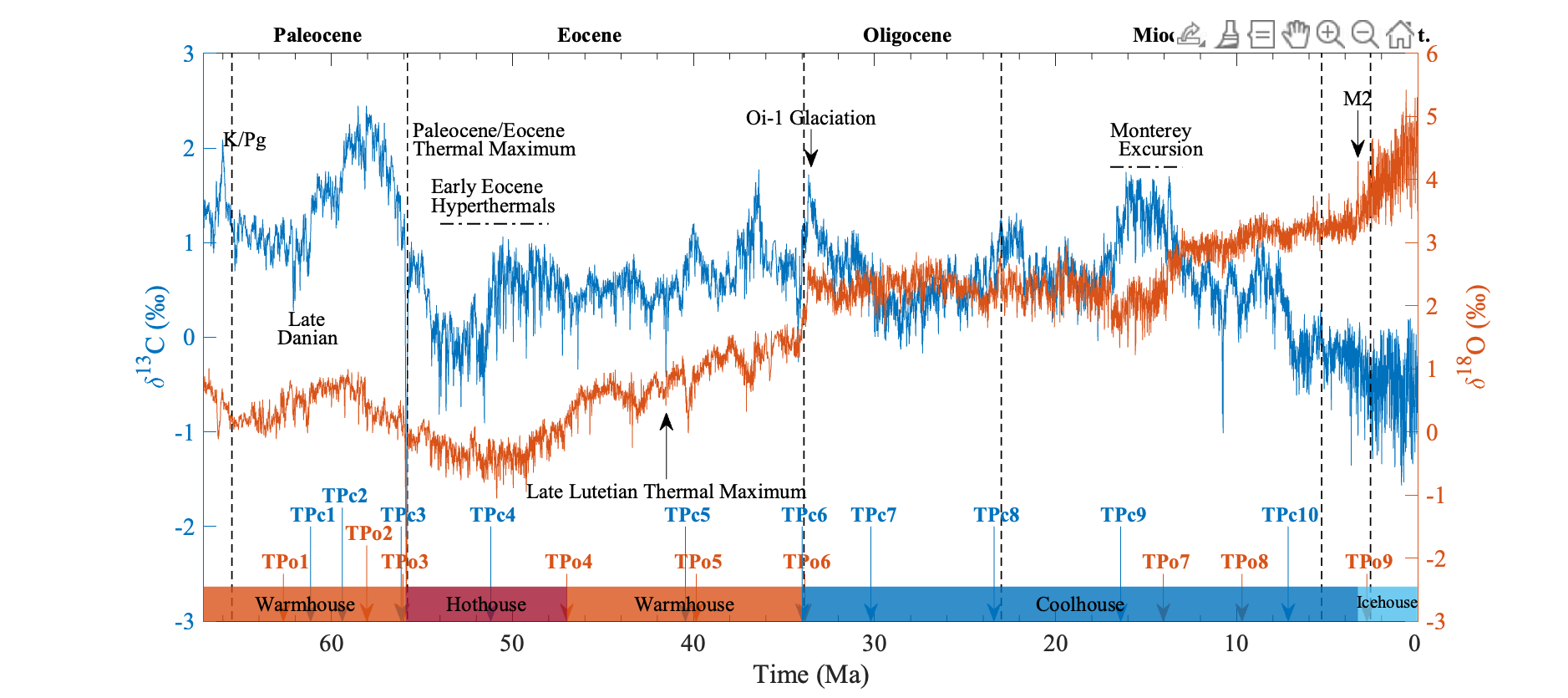}
    \caption{Time series of the CENOGRID $\delta^{13}$C (blue) and $\delta^{18}$O (red) records through the last 66 My. Vertical dashed lines mark the different geological epochs, while specific events are reported when they occurred. Arrows marks the Tipping Points (TPs) identified by Rousseau et al.~\cite{Rousseau23} with a uni-variate approach (see Supplementary Information).}
    \label{fig1}
\end{figure}

Earth’s climatic history has been reconstructed using sediment archives from both marine and terrestrial environments. In particular, the development of high-resolution deep-sea capturing oxygen isotopes ($\delta^{18}$O) and carbon isotopes ($\delta^{13}$C) records has since the 1970s \cite{Savin75,Kennett75} greatly enhanced understanding of past climate trends, cyclic variations, rates of change, and transient events throughout the Cenozoic era (66 My ago to present). However, the compilations have suffered limitations in accurately documenting the full range and detailed characteristics of Cenozoic climate variability, due to gaps and insufficient age control and temporal resolution, especially for the period before 34 My ago. 
A study \cite{Westerhold20} addressed these challenges by utilizing sediment archives obtained by the International Ocean Discovery Program (IODP) and its predecessor programs (DSDP, ODP) to compile and analyze a comprehensive new composite record of carbon and oxygen isotopes in deep-sea benthic foraminifera that was precisely tuned to astronomical cycles. The new climate reference curve, CENOGRID (CENOzoic Global Reference benthic foraminiferal carbon and oxygen Isotope Dataset) \cite{Westerhold20}, provides high-resolution coverage of the past 66 My to detect long-term Cenozoic climate variability (Figure \ref{fig1}). 

Employing statistical recurrence analysis (RA) on the CENOGRID record, Westerhold et al.~\cite{Westerhold20} identified four key climatic regimes in the Cenozoic era: Hothouse, Warmhouse, Coolhouse, and Icehouse, each demonstrating unique and statistically significant dynamics~\cite{Westerhold20}. The Warmhouse and Hothouse states prevailed from approximately 66 million years ago (at the Cretaceous/Paleogene boundary) to about 34 million years ago (at the Eocene-Oligocene Transition). During these periods, temperatures exceeded present-day levels by more than 5°C and 10°C, respectively. Notably, the Hothouse climate state witnessed transient warming events known as hyperthermals, marked by concurrent negative excursions in $\delta^{13}$C and $\delta^{18}$O, indicating substantial carbon release into the climate system and global warming~\cite{Lourens05,Nicolo07,Zachos10,Gutjahr17,Galeotti16}. The transition from the Warmhouse to the Coolhouse state occurred during the Eocene-Oligocene Transition, accompanied by a significant temperature drop and the establishment of a semi-permanent Antarctic Ice Sheet~\cite{Coxall05,Scher11,Galeotti16,Barr22,Rohling22}. The Coolhouse state extended from approximately 34 million years ago to 3.3 million years ago and comprised two phases, separated around 14 million years ago by a $\delta^{18}$O increase and transient $\delta^{13}$C rise in the deep ocean, signaling rapid Antarctic Ice Sheet expansion~\cite{Flower94,Raitzsch21}.
The Icehouse state, characterized by the fluctuation of ice sheets in the Northern Hemisphere, was fully established during the Pliocene-Pleistocene transition~\cite{Bailey13} and continues until the present. 

The Cenozoic climate has experienced several transitions \cite{Zachos2001,Westerhold20} associated with tipping points \cite{Lenton2008} of the Earth system. In \cite{Rousseau23} such transitions have been characterized, by combining recurrence analysis of the individual time series \cite{Marwan2007,Bagniewski2023} with a multi-variate analysis based on the quasi-potential theory \cite{Lucarini2020,Margazoglou2021}. In addition of these critical transitions between the four macroclusters of climate variability mentioned above, the analysis identified several other occurrences of tipping behaviour \cite{Rousseau23}. 

\begin{figure}[!h]
    \includegraphics[width=16cm]{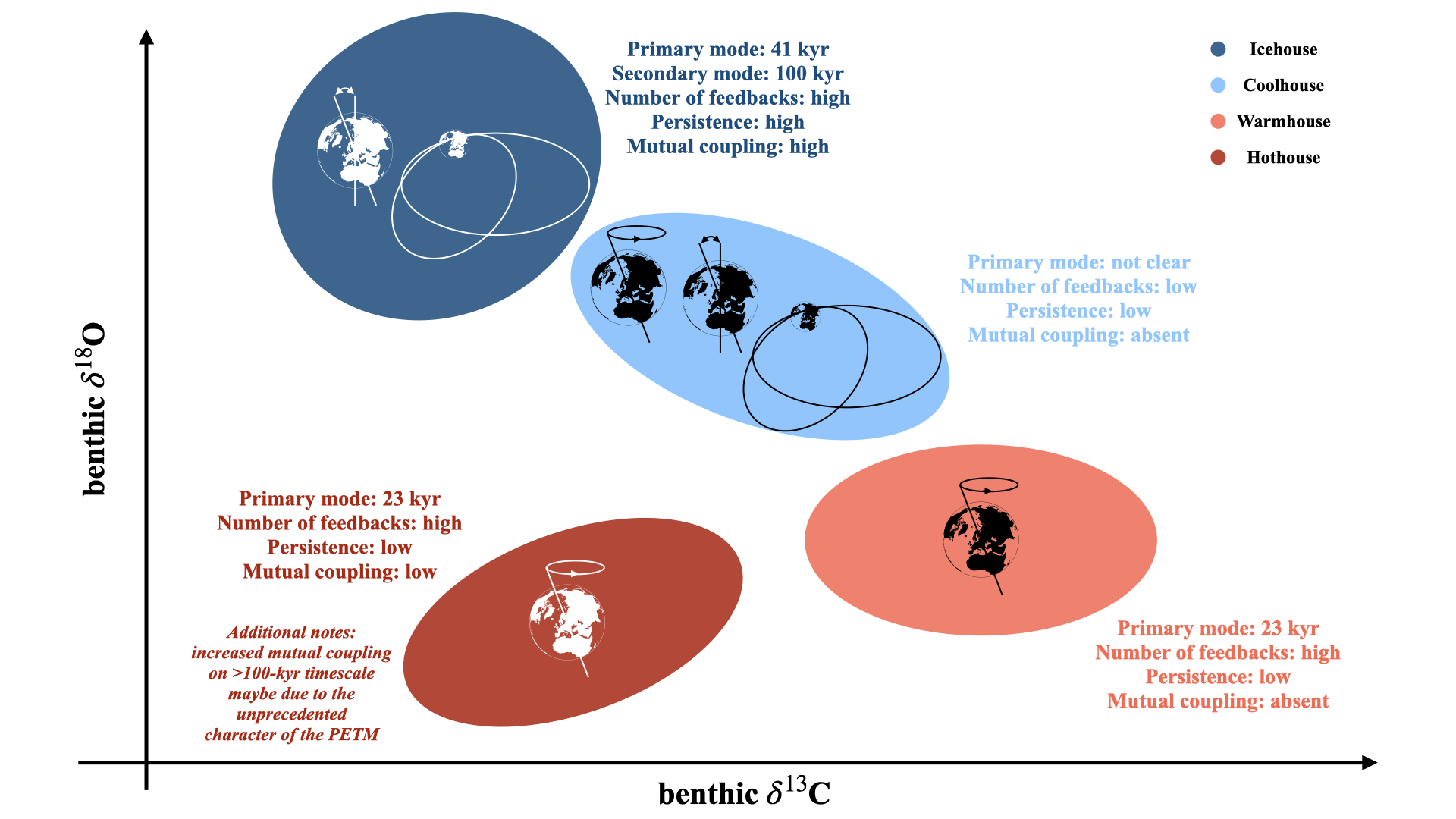}
    \caption{\textbf{Sketch of the main features of the four climate states.} The four climate states are ranked in terms of number and timescales of effective degrees of freedom, stability/predictability, and $\delta^{13}$C-$\delta^{18}$O coupling at multi-millennial timescales.}
    \label{fig0}
\end{figure}

Here we want to complement the previous investigations of the CENOGRID data by applying a novel multiscale and bivariate dimensional analysis of the CENOGRID record~\cite{Alberti23a,Alberti23b} to characterize the four climate states in terms of number of effective degrees of freedom and associated timescales, stability/predictability in the record with a focus on the critical transitions, and $\delta^{13}$C-$\delta^{18}$O coupling at multi-millennial timescales (see schematic in Figure \ref{fig0}).

So far, this multiscale, bivariate dimensional analysis has proved great skills in other different applications \cite{Alberti23a,Alberti23b}. The analysis extracts intrinsic scale-dependent components from the CENOGRID $\delta^{13}$C and $\delta^{18}$O records via Empirical Mode Decomposition (EMD) \cite{Huang98} and determines, across timescales ($\tau$), three key parameters \cite{Lucarini_2012jsp,lucarini2016extremes,Faranda17}: (1) instantaneous dimension $d$, which provides an estimate of the number of effective degrees of freedom of the system (linked to the presence of stronger positive feedback mechanisms acting on the system preferentially at the timescale $\tau$ leads to higher values of $d$); (2) extremal index $\theta$, which quantifies the stability with respect to perturbations of the system; i.e., its intrinsic persistence, where $\theta$ close to 0 (1) means that the system is more (less) persistent; and (3) co-recurrence ratio $\alpha$, which quantifies the mutual coupling between  two proxies, with $\alpha$ close to 0 (1) meaning stronger (weaker) coupling. It is well known that, depending on the timescale of interest, the same system might exhibit different stability properties, depending as to whether positive or negative feedbacks dominate \cite{Arnscheidt2022}. We remark that these indicators should be interpreted in relative rather than absolute terms. Rather than taking at face value the obtained estimate of $d$, we proceed as follows. If, e.g., state (a) features a larger value of $d$ than state (b), then we conclude that the number of effective degrees of freedom of state (a) is larger.
The key aspects of the EMD, an in-depth description of the uni-variate and bi-variate parameters, and our scale-dependent procedure are reported in the Methods section.

\section*{Multiscale analysis of the CENOGRID dataset}

%
\begin{figure}[!h]
    \includegraphics[width=16cm]{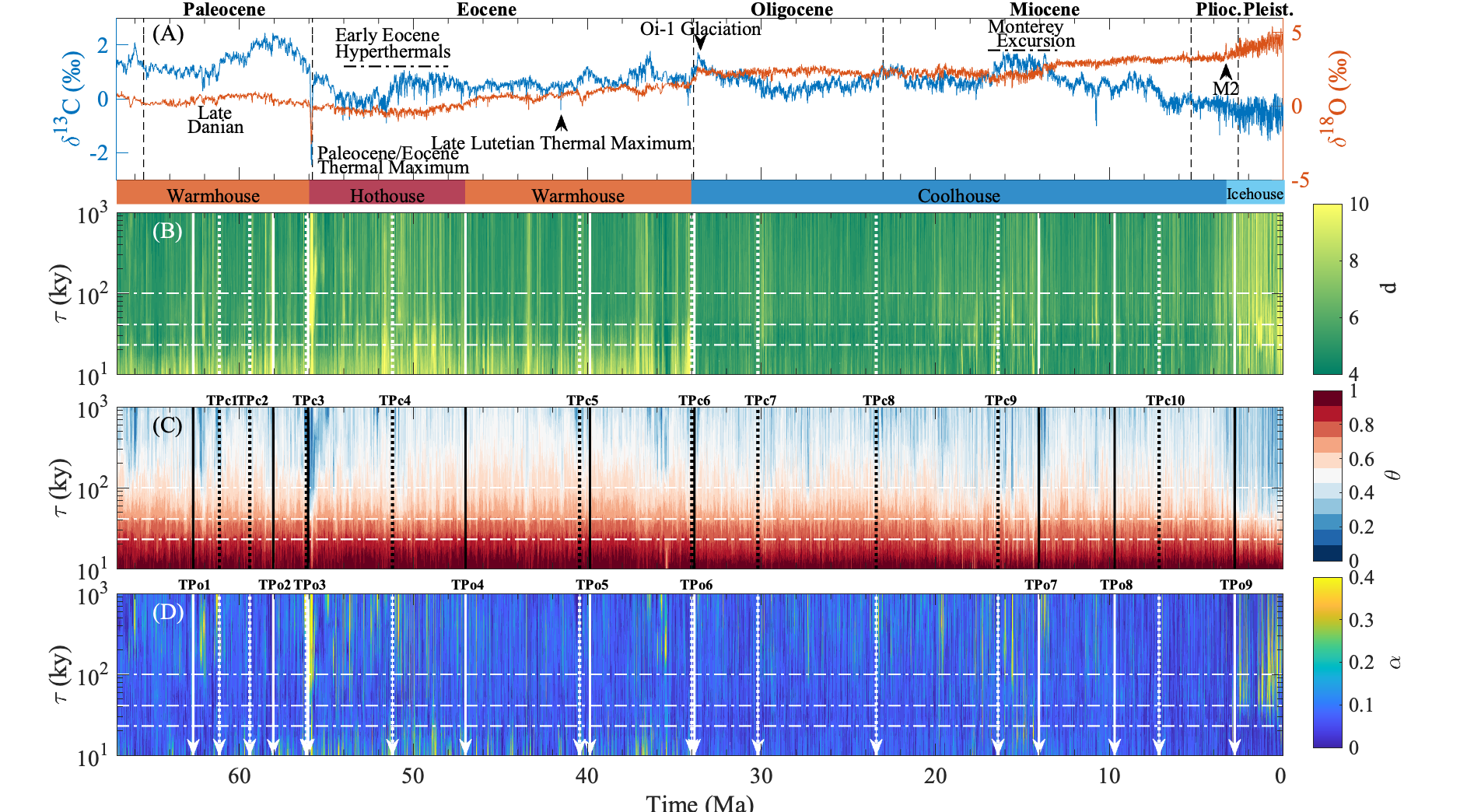}
    \caption{\textbf{Multiscale bivariate analysis during the last 66 Ma.} (A) Time series of the CENOGRID paleoclimate records of $\delta^{13}C$ (blue) and $\delta^{18}O$ (red) during the last 66 Ma. Bi-variate multi-scale metrics: (B) instantaneous dimension $d$, (C) extremal index $\theta$, and (D) co-recurrence index $\alpha$. Vertical dashed lines in panel (A) mark the different geological epochs, while specific events are reported when they occurred. Vertical continuous and dotted lines in panels (B)-(D) identify the Tipping Points (TPs) identify by Rousseau et al.~\cite{Rousseau23} with a uni-variate approach (see Supplementary Information) for $\delta^{18}$O and $\delta^{13}$C, respectively. The horizontal dashed-dotted white lines refer to Milankovitch scales of precession ($\sim23$ ky), obliquity ($\sim41$ ky), and eccentricity ($\sim100$ ky), respectively.}
    \label{fig2}
\end{figure}

The behavior of our multiscale bivariate metrics (Figure \ref{fig2}) highlights the properties of the climate variability recorded in the CENOGRID dataset. The instantaneous dimension $d$ (Figure \ref{fig2}A) clearly indicates a lower number of effective degrees of freedom during the Coolhouse than during the other climate states, except for an increase during specific events such as the Monterey positive carbon isotope excursion between 16.9 and 13.5 My ago \cite{Sosdian20} that lacks dominant expression of any specific orbital period. Conversely, there is a net difference when comparing warmer and cooler climate states in terms of the dominant scales at which climate responses to orbital forcing(s) are observed. By looking at the timescale-dependent estimate of $d$, one observes that whilst the Warmhouse and the Hothouse states are dominated by active positive feedbacks at short orbital timescales (precession mainly), the Icehouse is dominated by active positive feedbacks at obliquity and eccentricity timescales. 
For warm climates this can be related to the moderate-to-high levels of carbon dioxide and the existence of a mix of tropical and temperate ecosystems. Conversely, for the Icehouse this can be related to the presence of enhanced polar ice sheets and glaciers and oceans, enhancing ice-ocean-atmosphere coupling processes, and associated greenhouse gas variability. In general, we observe substantial short-term mutual coupling (increased $\alpha$; Figure \ref{fig2}D) between $\delta^{13}$C and $\delta^{18}$O during warm climate states, which no longer occurs during cold climate states, together with temporarily strongly enhanced coupling during specific intervals such as the Late Danian, Early Eocene Hyperthermals, and the middle Miocene Monterey event. During the Icehouse, starting from the M2 glaciation, prolonged mutual coupling is found at obliquity and eccentricity timescales, which did not occur before. This is related to the repeated glacial-interglacial variations, whereby a strong coupling is established between the average surface temperature and the intensity of the carbon cycle. Furthermore, while all climate states are characterized by non-persistent behavior (high $\theta$; Figure \ref{fig2}C) at timescales shorter than the obliquity period, increased persistence (lower $\theta$) is observed at obliquity and eccentricity timescales during the Icehouse climate state, marking the current climate state as the one with most persistent responses at orbital 41- and 100-ky timescales out of the entire past 66 My, related to the glacial-interglacial cycles.

Finally, our analysis highlights the exceptional nature, even among other hyperthermals, of the Paleocene Eocene Thermal Maximum (PETM), where $d$ is high and almost constant across all timescales and lacks association with any particular orbital timescale, while coupling is high at all timescales (high $\alpha$; Figure \ref{fig2}C). In contrast, the Eocene-Oligocene Transition (EOT, $\sim$34 Ma) seems to be primarily influenced by processes at precession timescales, marking the end of the precession-dominated Hot-/Warm-house period. Both these transitions were identified recently as key abrupt transitions (Tipping Points, TPs) associated with major regime shifts that separate clusters of climate variability \cite{Rousseau23}. We confirm this using a uni-variate framework as in Ref. \cite{Rousseau23} (Figures S1-S2). 

\section*{Dynamical Features of the Four Macroclusters of CLimate Variability across Timescales}

\begin{figure}[!h]
    \includegraphics[width=16cm]{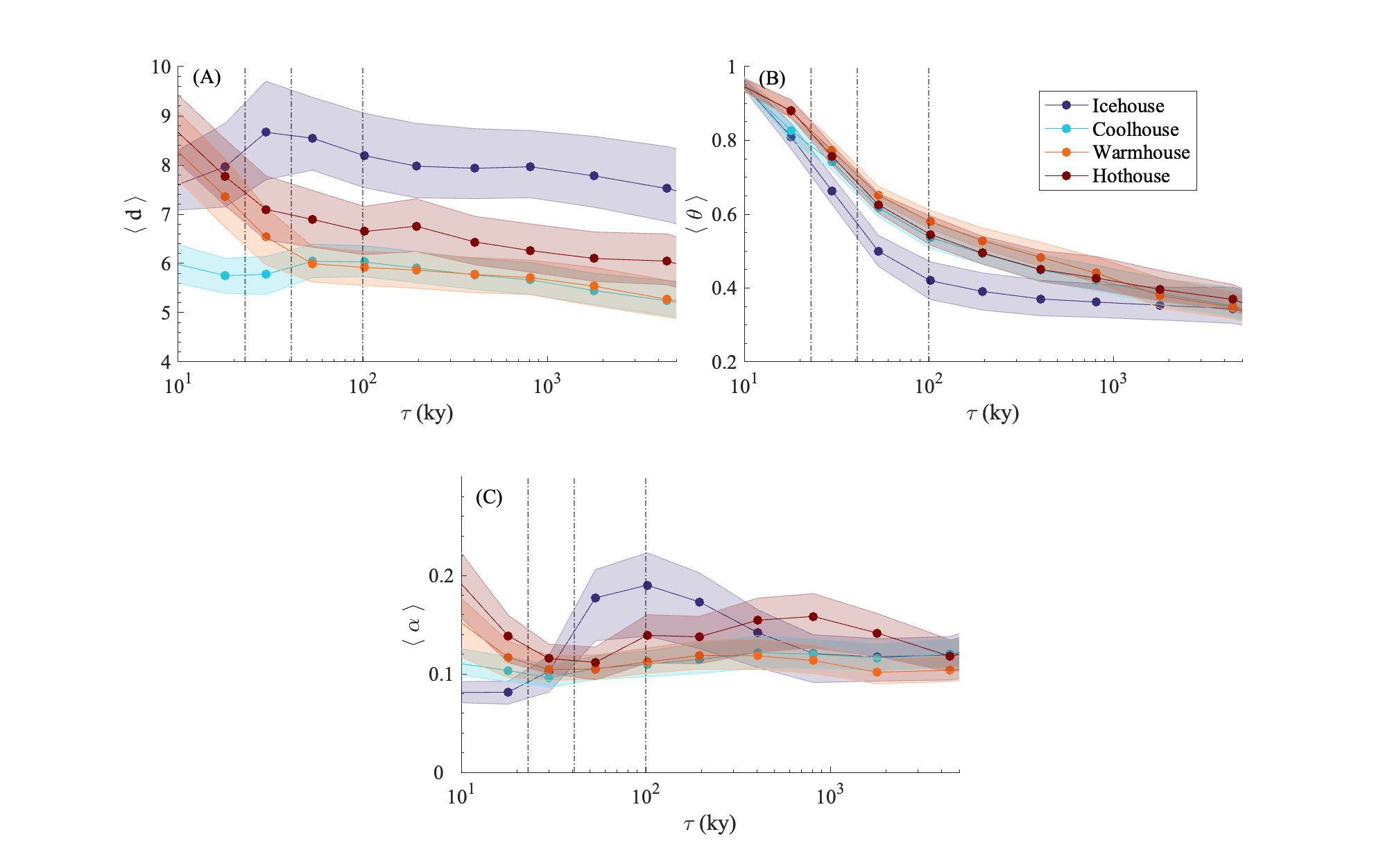}
    \caption{\textbf{Time-averaged multi-scale bi-variate statistics during the four climate states identified by \cite{Westerhold20}.} Temporal averages of the bi-variate multi-scale metrics: (A) instantaneous dimension $\langle d \rangle$, (B) extremal index $\theta$, and (C) co-recurrence ratio $\alpha$. The error bars refers to the inter-quartile range. The vertical dashed-dotted black lines refer to Milankovitch scales of precession ($\sim23$ ky), obliquity ($\sim41$ ky), and eccentricity ($\sim100$ ky), respectively.}
    \label{fig3}
\end{figure}

To facilitate interpretation of our results, we investigate the average values of $d$, $\theta$, and $\alpha$ in the timescale domain ($\tau$) for the four climate states (Fig. \ref{fig3}). 

The Hothouse climate state is characterized by $d$ decrease with increasing $\tau$ with a bit of a plateau around the 100-ky eccentricity period, which indicates the largest number of degrees of freedom in the climate response at around the 23-ky precession timescale. A similarly decreasing trend with $\tau$ is visible for $\theta$, which suggests increased climate stability at the dominant (precession) period of variability. Finally, $\alpha$ is clearly high at precession timescales, as well as at timescales larger than the eccentricity period. The former can be related to responses to precession variations, while the latter is mainly due to the PETM signature.
The Warmhouse state is also characterized by decreasing $d$ and $\theta$ with increasing $\tau$, while $d$ values are lower and $\theta$ values similar, relative to the Hothouse. As opposed to the Hothouse, the Warmhouse is characterized by approximately constant $\alpha$ for all timescales at around 0.1, which indicates low mutual coupling between $\delta^{13}$C-$\delta^{18}$O.

During the Coolhouse state, a different behavior is observed, characterized by absence of a clearly dominant timescale for $d$. 
Yet, $\theta$ and $\alpha$ decrease with $\tau$ in a similar manner as observed in the warm climate states, and values are also similar to those in the warm states. This decrease in all metrics reflects relatively stable conditions during the Coolhouse state. Finally, results differ completely for the Icehouse state, with a high overall number and slight timescale-dependence for $d$, which peaks at around the 41-ky obliquity timescale. This suggests an increased number of degrees of freedom involved in climate responses at all timescales, relative to the prior warm climate states and the Coolhouse state. This reflects the increased ice-ocean-atmosphere interactions and associated ecosystem variations. Meanwhile, extremal index $\theta$, while still characterized by a decrease with $\tau$, shows considerably lower values (i.e., greater stability) compared to the previous climate states. Additionally, $\alpha$ is once again highly timescale dependent, but now peaks at 41- and 100-ky timescales, reaching higher values than previously observed. 

Overall, our analysis implies that warmer and colder climates respond substantially differently to orbital forcing. Responses during warm climates are dominated by precession timescale variations, whereas cold climates appear to be driven mainly by responses on obliquity and eccentricity timescales. 

\begin{figure}[!h]
    \includegraphics[width=16cm]{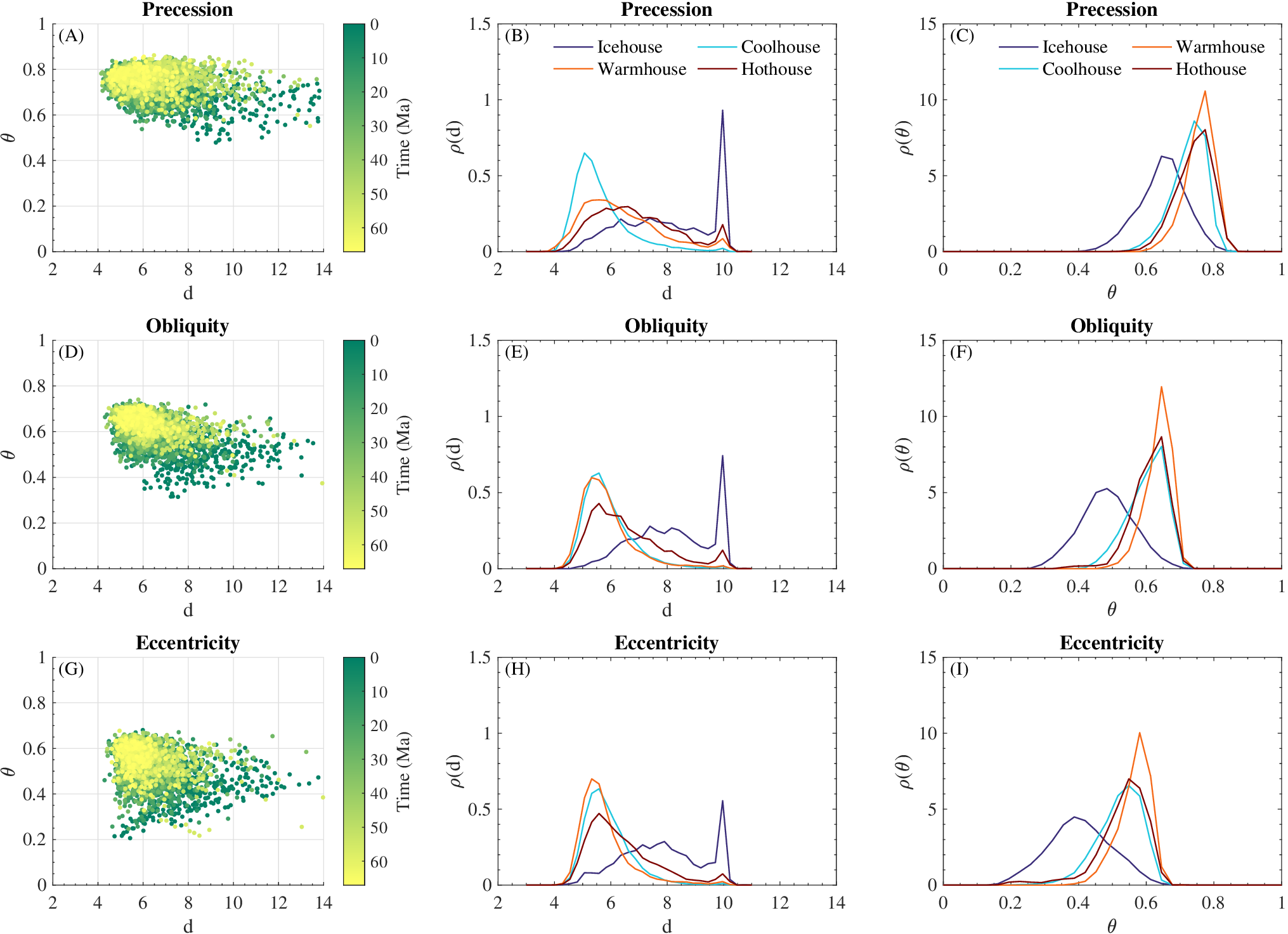}
    \caption{\textbf{Multiscale bivariate scatter-plots of the metrics at Milankovitch scales.} $d-\theta$ scatter plots colored by time instants at the three Milankovitch scales of precession (A), obliquity (D), and eccentricity (G), respectively. The distribution of the instantaneous dimension $d$ over the four different climate states identified by \cite{Westerhold20} at the three Milankovitch scales of precession (B), obliquity (E), and eccentricity (H), respectively. The distribution of the extremal index $\theta$ over the four different climate states identified by \cite{Westerhold20} at the three Milankovitch scales of precession (C), obliquity (F), and eccentricity (I), respectively.}
    \label{fig4}
\end{figure}

To further inspect the climate response at orbital timescales, we investigate the 2-D ($d$, $\theta$) parameter-space behavior and the probability distribution functions (pdfs) in the four climate states (Fig. \ref{fig4}), as well as in different geological epochs (Supplementary Figure S3) and in a uni-variate framework between two consecutive TPs identified by Rousseau et al. \cite{Rousseau23} (Supplementary Figures S4, S5).

Moving across orbital timescales (Fig. \ref{fig4} (A), (D), (G)) there is a transition toward a different portion of the 2-D parameter-space when approaching the Icehouse (starting from the late Coolhouse). Furthermore, a wider spread in the range of values of both $d$ and $\theta$ is observed at the eccentricity timescale, while a narrower $\theta$-range occurs at the precession scale. The larger $d$ range is seen during the late Coolhouse and the entire Icehouse; conversely, a confined region in the $d$-$\theta$ space is associated with both Hothouse and Warmhouse states. This is confirmed by the $d$ and $\theta$ pdfs for the four climate states (Fig. \ref{fig4} (B)-(C), (E)-(F), (H)-(I)). The uniqueness of the Icehouse state is evidenced by the $d$ distribution across orbital timescales which peaks at $\sim$10 for all Milankovitch cycles. The $d$ distribution is also almost similar at all orbital timescales during the Coolhouse, albeit with a peak at lower values ($\sim$5). Conversely, the two warm states are characterized by different $d$ distributions across orbital timescales: a wider $d$ spread occurs at the precession timescale, and more peaked distributions at obliquity and eccentricity timescales, which also shift toward lower values ($\sim$5). The $\theta$ pdfs confirm the uniqueness of the Icehouse state: for all orbital timescales, $\theta$ is lower during the Icehouse and the shape of the pdf is completely different relative to those of the other states, which turn are very similar to each other. 

The unique nature of climate during the last $\sim$5 My is again evident from inspection of the dynamical properties of the Cenozoic climates compared with climates in different geological epochs with a similar bivariate approach (Supplementary Figure S3), as well as between two consecutive TPs of Rousseau et al. \cite{Rousseau23} using a uni-variate framework (Supplementary Figures S4, S5). Our stronger result is that the Icehouse, or equivalently the Plio-Pleistocene (or, equivalently, the period starting from the last TPs of Ref. \cite{Rousseau23}), represents an unprecedented state in Earth’s Cenozoic climate system. It is characterized by co-existence of widespread variability on two primary orbital timescales, through a large 
number of effective degrees of freedom and with enhanced mutual coupling between atmosphere and ocean. In contrast, warmer climate states (Hothouse and Warmhouse) are characterized by a response at precession timescales, with relatively low average and instantaneous numbers of effective degrees of freedom and low (Hothouse) to absent (Warmhouse) mutual coupling between $\delta^{13}$C and $\delta^{18}$O, apart from a high level of coupling during the exceptional PETM (mainly at timescales larger than the eccentricity period). The Coolhouse state is markedly different from both the warmer climate states and the Icehouse state, with low effective degrees of freedom and coupling across timescales that lack a primary mode of variability for any orbital cycle. 

\section*{Conclusions and Outlook}

The Cenozoic era, spanning the last 66 million years, has witnessed significant changes in Earth’s climate \cite{Westerhold20}, including a large number of diverse critical transitions \cite{Rousseau23}. Understanding the physical processes involved in this variability is crucial for interpreting paleoclimate data and projecting future scenarios \cite{IPCC2022_Mitigation}. Our analysis clearly highlights the crucial impact of polar ice sheet formation and evolution in regulating global climate, and feedback mechanisms have been critical to shaping these ice sheets. For example, polar ice sheet growth causes enhanced reflection of sunlight back into space, which causes further cooling that, in turn, fosters further ice growth (the positive ice-albedo feedback). Moreover, ice sheet waxing and waning is crucial in ocean-atmosphere coupling, which is at the basis of heat transport across the globe and thermal regulation of climate. And ice sheet fluctuations also affected vegetation-zone displacements, which further affect surface albedo and, thus, the energy balance of climate \cite{Alberti15,Rombouts15}. 

We show here that the critical transitions identified  in \cite{Rousseau23} are accompanied by anomalously large value in the number of effective degrees of freedom and anomalously low values for the extremal index, which suggests dominant impacts of positive feedbacks. The instability in the extent of ice sheets during the Icehouse states is also associated with the anomalously high value of $d$ found in this period. 
In turn, 
our finding of an increased persistence (stability) of the Icehouse climate state agrees with a relatively stable Antarctic ice cover over extended periods despite varying climate conditions. 


The climate responses to orbital forcing -which influences mainly the spatial and temporal distributions, and (weakly) the total amount of solar radiation reaching Earth’s surface - consist of a complex interplay of positive and negative feedbacks. wW find that positive feedbacks are more effective on specific orbital periods under different climate states. We detect that they acted on a dominant 23-ky (precession) timescale during Hothouse and Warmhouse climate states, mixed 41-ky (obliquity) and 100-ky (eccentricity) timescales during the Icehouse state, and that there is a remarkable lack of dominant timescales during the Coolhouse state. 
The detection of the primary variability timescales provides fundamental insights into the drivers of long-term climate variability and confirms the multiscale nature of climate variability \cite{Ghil20,Anna2021}. This understanding is essential for understanding their significance in the context of future climate changes, and for assessing the capability of (paleo-)climate models to adequately replicate climate states and critical transitions between and within them.


Finally, we demonstrate the robustness and the power of our analysis for objective identification of abrupt climate shifts and extremes using high-resolution records, such as the highly (20-y) resolved Greenland NGRIP $\delta^{18}$O record \cite{Andersen04}. This analysis identifies Dansgaard-Oeschger (D-O) events as rapid transitions within a single state \cite{Alberti14}, in contrast to major transitions between states observed above in CENOGRID. Our analysis reveals a similar nature between D-O fluctuations and PETM or hyperthermals, being both extreme fluctuations featuring high $d$ and low $\theta$. Specifically, our results indicate that the D-O fluctuations, the PETM, and the hyperthermals appear as transitions that involve many feedbacks (high $d$) in a generally stable condition (low $\theta$); hence, they are coherent system shifts, instead of transitions into different climate states. We observe a deviation only for the Holocene (lower $d$ and higher $\theta$) relative to the glacial cycle, which we still cannot associate with a climate state change (as observed in CENOGRID), but which marks the interglacial as being at a different end of the spectrum than the interstadial/stadial fluctuations of MIS3. Thus, use of our method on high-resolution records can offer perspectives on the role of the different mechanisms underpinning climate shifts and transitions at different timescales. Our next question will be whether anthropogenic forcing is driving a climate state transition out of the Icehouse.


\paragraph*{Funding:} V.L. acknowledges the support received from the EPSRC project EP/T018178/1 and from the EU Horizon 2020 project TiPES (Grant no. 820970).

\paragraph*{Author contributions:} T.A. and F.F. conceived the study (conceptualization) and wrote the first draft of the manuscript (writing – original draft). All authors interpreted, edited, and reviewed the manuscript (writing – review and editing).
\paragraph*{Competing interests:} The authors declare no competing interests.
\paragraph*{Data and materials availability:} All data are available open access in electronic form at the PANGAEA data repository (\url{https://doi.org/10.1594/PANGAEA.917503}). Numerical resources for uni-variate and bi-variate dynamical system metrics are freely available at \url{https://www.davide-faranda.com/scripts}. 

\bibliography{Albertietal}

\bibliographystyle{Science}

\end{document}